\documentstyle[twoside,epsfig]{article}
\catcode`\@=11
\long\def\@makefntext#1{
\protect\noindent \hbox to 3.2pt {\hskip-.9pt  
$^{{\eightrm\@thefnmark}}$\hfil}#1\hfill}		%CAN BE USED 

\def\thefootnote{\fnsymbol{footnote}}
\def\@makefnmark{\hbox to 0pt{$^{\@thefnmark}$\hss}}	%ORIGINAL 
	
\def\ps@myheadings{\let\@mkboth\@gobbletwo
\def\@oddhead{\hbox{}
\rightmark\hfil\eightrm\thepage}   
\def\@oddfoot{}\def\@evenhead{\eightrm\thepage\hfil
\leftmark\hbox{}}\def\@evenfoot{}
\def\sectionmark##1{}\def\subsectionmark##1{}}

\oddsidemargin=\evensidemargin
\renewcommand{\thefootnote}{\fnsymbol{footnote}}

\newcounter{sectionc}\newcounter{subsectionc}\newcounter{subsubsectionc}
\renewcommand{\section}[1] {\vspace{12pt}\addtocounter{sectionc}{1} 
\setcounter{subsectionc}{0}\setcounter{subsubsectionc}{0}\noindent 
	{\tenbf\thesectionc. #1}\par\vspace{5pt}}
\renewcommand{\subsection}[1] {\vspace{12pt}\addtocounter{subsectionc}{1} 
	\setcounter{subsubsectionc}{0}\noindent 
	{\bf\thesectionc.\thesubsectionc. {\kern1pt \bfit #1}}\par\vspace{5pt}}
\renewcommand{\subsubsection}[1] {\vspace{12pt}\addtocounter{subsubsectionc}{1}
	\noindent{\tenrm\thesectionc.\thesubsectionc.\thesubsubsectionc.
	{\kern1pt \tenit #1}}\par\vspace{5pt}}

\newcounter{appendixc}
\newcounter{subappendixc}[appendixc]
\newcounter{subsubappendixc}[subappendixc]
\renewcommand{\thesubappendixc}{\Alph{appendixc}.\arabic{subappendixc}}
\renewcommand{\thesubsubappendixc}
	{\Alph{appendixc}.\arabic{subappendixc}.\arabic{subsubappendixc}}

\renewcommand{\appendix}[1] {\vspace{12pt}
        \refstepcounter{appendixc}
        \setcounter{figure}{0}
        \setcounter{table}{0}
        \setcounter{lemma}{0}
        \setcounter{theorem}{0}
        \setcounter{corollary}{0}
        \setcounter{definition}{0}
        \setcounter{equation}{0}
        \renewcommand{\thefigure}{\Alph{appendixc}.\arabic{figure}}
        \renewcommand{\thetable}{\Alph{appendixc}.\arabic{table}}
        \renewcommand{\theappendixc}{\Alph{appendixc}}
        \renewcommand{\thelemma}{\Alph{appendixc}.\arabic{lemma}}
        \renewcommand{\thetheorem}{\Alph{appendixc}.\arabic{theorem}}
        \renewcommand{\thedefinition}{\Alph{appendixc}.\arabic{definition}}
        \renewcommand{\thecorollary}{\Alph{appendixc}.\arabic{corollary}}
        \renewcommand{\theequation}{\Alph{appendixc}.\arabic{equation}}
        \noindent{\tenbf Appendix \theappendixc #1}\par\vspace{5pt}}
\newcommand{\subappendix}[1] {\vspace{12pt}
        \refstepcounter{subappendixc}
        \noindent{\bf Appendix \thesubappendixc. {\kern1pt \bfit #1}}
	\par\vspace{5pt}}
\newcommand{\subsubappendix}[1] {\vspace{12pt}
        \refstepcounter{subsubappendixc}
        \noindent{\rm Appendix \thesubsubappendixc. {\kern1pt \tenit #1}}
	\par\vspace{5pt}}

\newcommand{\textlineskip}{\baselineskip=13pt}
\newcommand{\smalllineskip}{\baselineskip=10pt}

\def\eightcirc{
\begin{picture}(0,0)
\put(4.4,1.8){\circle{6.5}}
\end{picture}}
\def\eightcopyright{\eightcirc\kern2.7pt\hbox{\eightrm c}} 

\newcommand{\copyrightheading}[1]
	{\vspace*{-2.5cm}\smalllineskip{\flushleft
	{\footnotesize International Journal of Modern Physics B, #1}\\
	{\footnotesize $\eightcopyright$\, World Scientific Publishing
	 Company}\\
	 }}

\def\abstracts#1#2#3{{
	\centering{\begin{minipage}{4.5in}\baselineskip=10pt\footnotesize
	\parindent=0pt #1\par 
	\parindent=15pt #2\par
	\parindent=15pt #3
	\end{minipage}}\par}} 

\renewenvironment{thebibliography}[1]			%ALL CHANGES DD 13/3/92
	{\frenchspacing
	 \ninerm\baselineskip=11pt
	 \begin{list}{\arabic{enumi}.}
	{\usecounter{enumi}\setlength{\parsep}{0pt}
	 \setlength{\leftmargin 12.7pt}{\rightmargin 0pt} %FOR 1--9 ITEMS
	 \setlength{\itemsep}{0pt} \settowidth
	{\labelwidth}{#1.}\sloppy}}{\end{list}}

\newcounter{itemlistc}
\newcounter{romanlistc}
\newcounter{alphlistc}
\newcounter{arabiclistc}

\newcommand{\fcaption}[1]{
        \refstepcounter{figure}
        \setbox\@tempboxa = \hbox{\footnotesize Fig.~\thefigure. #1}
        \ifdim \wd\@tempboxa > 5in
           {\begin{center}
        \parbox{5in}{\footnotesize\smalllineskip Fig.~\thefigure. #1}
            \end{center}}
        \else
             {\begin{center}
             {\footnotesize Fig.~\thefigure. #1}
              \end{center}}
        \fi}

\newcommand{\tcaption}[1]{
        \refstepcounter{table}
        \setbox\@tempboxa = \hbox{\footnotesize Table~\thetable. #1}
        \ifdim \wd\@tempboxa > 5in
           {\begin{center}
        \parbox{5in}{\footnotesize\smalllineskip Table~\thetable. #1}
            \end{center}}
        \else
             {\begin{center}
             {\footnotesize Table~\thetable. #1}
              \end{center}}
        \fi}

\def\@citex[#1]#2{\if@filesw\immediate\write\@auxout
	{\string\citation{#2}}\fi
\def\@citea{}\@cite{\@for\@citeb:=#2\do
	{\@citea\def\@citea{,}\@ifundefined
	{b@\@citeb}{{\bf ?}\@warning
	{Citation `\@citeb' on page \thepage \space undefined}}
	{\csname b@\@citeb\endcsname}}}{#1}}

\newif\if@cghi
\def\cite{\@cghitrue\@ifnextchar [{\@tempswatrue
	\@citex}{\@tempswafalse\@citex[]}}
\def\citelow{\@cghifalse\@ifnextchar [{\@tempswatrue
	\@citex}{\@tempswafalse\@citex[]}}
\def\@cite#1#2{{$\null^{#1}$\if@tempswa\typeout
	{IJCGA warning: optional citation argument 
	ignored: `#2'} \fi}}

\def\pmb#1{\setbox0=\hbox{#1}
	\kern-.025em\copy0\kern-\wd0
	\kern.05em\copy0\kern-\wd0
	\kern-.025em\raise.0433em\box0}

\def\fnt#1#2{\footnotetext{\kern-.3em
	{$^{\mbox{\scriptsize #1}}$}{#2}}}

\def\fpage#1{\begingroup
\voffset=.3in
\thispagestyle{empty}\begin{table}[b]\centerline{\footnotesize #1}
	\end{table}\endgroup}

\def\runninghead#1#2{\pagestyle{myheadings}
\markboth{{\protect\footnotesize\it{\quad #1}}\hfill}
{\hfill{\protect\footnotesize\it{#2\quad}}}}
\font\tenrm=cmr10
\font\tenit=cmti10 
\font\tenbf=cmbx10
\font\bfit=cmbxti10 at 10pt
\font\ninerm=cmr9

\font\eightrm=cmr8

\def\qed{\hbox{${\vcenter{\vbox{			%HOLLOW SQUARE
   \hrule height 0.4pt\hbox{\vrule width 0.4pt height 6pt
   \kern5pt\vrule width 0.4pt}\hrule height 0.4pt}}}$}}

\renewcommand{\thefootnote}{\fnsymbol{footnote}}	%USE SYMBOLIC FOOTNOTE

\def\bsc{{\sc a\kern-6.4pt\sc a\kern-6.4pt\sc a}}	%LATEX LOGO
\def\bflatex{\bf L\kern-.30em\raise.3ex\hbox{\bsc}\kern-.14em 
T\kern-.1667em\lower.7ex\hbox{E}\kern-.125em X} 

\begin{document}

\runninghead{Forward electron-phonon scattering in normal and superconducting state} {Forward electron-phonon scattering in normal and superconducting state}

\normalsize\textlineskip
\thispagestyle{empty}
\setcounter{page}{1}

\copyrightheading{}			%{Vol. 0, No. 0 (1993) 000---000}

\vspace*{0.88truein}

\fpage{1}
\centerline{\bf FORWARD ELECTRON-PHONON  SCATTERING IN NORMAL }
\vspace*{0.035truein}
\centerline{\bf AND SUPERCONDUCTING STATES}
\vspace*{0.37truein}
\centerline{\footnotesize O.V.DOLGOV}
\vspace*{0.015truein}
\centerline{\footnotesize \it Institut f\"ur Theoretische Physik, Universit\"at T\"ubingen} 
\baselineskip=10pt
\centerline{\footnotesize\it T\"ubingen, Germany} 
\vspace*{10pt}
\centerline{\footnotesize O.V.DANYLENKO}
\vspace*{0.015truein}
\centerline{\footnotesize \it P. N. Lebedev Physical Institute, Moscow, Russia } 
\vspace*{10pt}
\centerline{\footnotesize M. L. KULI\'C }
\vspace*{0.015truein}
\centerline{\footnotesize \it Centre de
Physique Th\'eorique et de Mod\'elisation, Universit\'e Bordeaux I, France } 
\vspace*{10pt}
\centerline{\footnotesize V. OUDOVENKO}
\vspace*{0.015truein}
\centerline{\footnotesize \it Joint Institute for Nuclear Research,  Dubna, Russia} 
\vspace*{0.21truein}
\abstracts{The sharp forward electron-phonon $(FEP)$ and impurity $(FIS)$
 scattering
change the normal and superconducting properties significantly. The pseudo-gap
like features are present in the density of states for $\omega <\Omega $,
where $\Omega $ is the phonon frequency. The superconducting critical
temperature $T_c$, due to the $FEP$ pairing, is linear with respect to the
electron-phonon coupling constant. 
The $FIS$ impurities are pair weakening
for $s-$ and $d-wave$ pairing.}{}{}

%\vspace*{10pt}
%\keywords{The contents of the keywords}

%\textlineskip			%) USE THIS MEASUREMENT WHEN THERE IS
%\vspace*{12pt}			%) NO SECTION HEADING

\vspace*{1pt}\textlineskip	%) USE THIS MEASUREMENT WHEN THERE IS
\section{Introduction}	%) A SECTION HEADING
%\vspace*{-0.5pt}
\noindent
 There is growing experimental evidence for
d-wave pairing in high-T$_{c}$ superconductors (HTS), which are seemingly in
contradiction with the standard phonon mechanism of pairing. 
However, some optic and tunneling measurements show features which can be
ascribed to rather strong electron-phonon (E-P) coupling.
The idea
of d-wave pairing in HTS oxides due to the renormalized (by strong
electronic correlations) E-P coupling has been put forward in Ref.[1]. 
It was shown\cite{Kulic} that for small hole
doping, strong Coulomb correlations renormalize the E-P interaction
giving rise to the strong forward (small-q) scattering peak.
This
renormalization means that each quasiparticle, due to the suppression of the
doubly occupancy on the same lattice, is surrounded by a giant correlation
hole. The pronounced forward E-P scattering can be also due to
the poor Coulomb screening in HTS oxides as
well as due to the large density of states near some points at the
Fermi surface \cite{Abrikosov}. 
In the following analysis we
assume an extreme case of the forward electron-phonon interaction (FEP
pairing) and of the forward nonmagnetic impurity scattering ($FIS$
impurities), i.e. that $\mid g_{scr}(\vec{q})\mid ^{2}\sim \delta (\vec{q})$
and $u_{scr}^{2}(\vec{q})\sim \delta (\vec{q})$, where $\delta (\vec{q})$ is
the Dirac delta-function, $g_{scr}(\vec{q})$ and $u_{scr}(\vec{q})$ are the
screened E-P and impurity potentials respectively. This approximation picks
up the main physics, and it is valid whenever the range $R$ of $\mid g_{scr}(
\vec{q})\mid ^{2}$ and $u_{scr}^{2}(\vec{q})$ fulfils the condition $R\gg
k_{F}^{-1}$, i.e. $q_{c}\ll k_{F}$. Moreover, it greatly
simplifies the structure of the Eliashberg equations by omitting integration
in $\vec{k}$-space. A similar approximation was successfuly used for the AF
spin-fluctuation mechanism of pairing, where the four peaks at $\vec{Q}=(\pm
\pi ,\pm \pi )$ in the spin-fluctuation density were replaced by delta-functions \cite{Kostur}.
\textheight=7.8truein
\setcounter{footnote}{0}
\renewcommand{\thefootnote}{\alph{footnote}}

\section{Eliashberg equations for FEP pairing and FIS impurities}
\noindent
The Eliashberg equations in the presence of the $FEP$ pairing potential $
\delta (\vec k)V_{ep}(\omega )$ and of the $FIS$
impurities $\delta (\vec k)u_{imp}^2$, where the latter is
treated first in the self-consistent Born approximation. The renormalization
function $Z$, the energy renormalization $\bar \xi$ 
and the superconducting order parameter 
$\Delta$ are solutions of the following
equations
\[
Z_n(\xi )=1+\frac T{\omega _n}\sum_mV_{eff}(n-m)\frac{\omega _mZ_m(\xi )}{
[\omega _mZ_m(\xi )]^2+\bar \xi _m^2(\xi )+[Z_m(\xi )\Delta _m(\xi )]^2}, 
\]
\[
\bar \xi _n(\xi )=\xi (\vec k)-T\sum_m\frac{V_{eff}(n-m)\bar \xi
_m(\xi) }{[\omega _mZ_m(\xi )]^2+\bar \xi _m^2(\xi )+[Z_m(\xi )\Delta _m(\xi )]^2}
\]
\begin{equation}  \label{eq.1}
Z_n(\xi )\Delta _n(\xi )=T\sum_m\frac{V_{eff}(n-m)Z_m(\xi )\Delta_m(\xi)}{[\omega _mZ_m(\xi )]^2+\bar \xi _m^2(\xi )+[Z_n(\xi )\Delta _m(\xi )]^2}.
\end{equation}

 The retardation 
({\it strong-coupling}) effects are considered in the Einstein model with
one phonon at a frequency $\Omega$. The solution of Eqs.(1) can be found
by iteration, where the first iteration step at $T=0$ and 
the analytical continuation give 
$\Re Z(\omega ,\xi )=1+\frac \lambda {2N(0)}\frac \Omega {(\Omega 
+|\xi|)^2-\omega ^2}$, and
$\Im Z(\omega ,\xi )=-\frac{\pi \lambda }{4N(0)}\delta (\omega \pm (\Omega 
+|\xi|))$. In contrast to the standard behaviour, where the spectral function 
$A(\omega ,\xi )$ has a
coherent peak at $\omega =\pm |\xi |$ and incoherent wings for $\omega
>\Omega ,$ the $FEP$ interaction leads to a rather broad peak at $\omega
=\pm |\xi |$ and very sharp peaks at $|\xi |\pm \Omega $. As a consequence
the density of states $N(\omega )$ obtained by the solution of Eqs.(1) (see Fig.1) shows a maximum at $\omega =0$, 
\begin{figure}[htbp]
\vspace*{13pt}
%\centerline{\vbox{\hrule width 5cm height0.001pt}}
%\vspace*{1.4truein}		%ORIGINAL SIZE=1.6TRUEIN x 100% - 0.2TRUEIN
\centerline{\psfig{figure=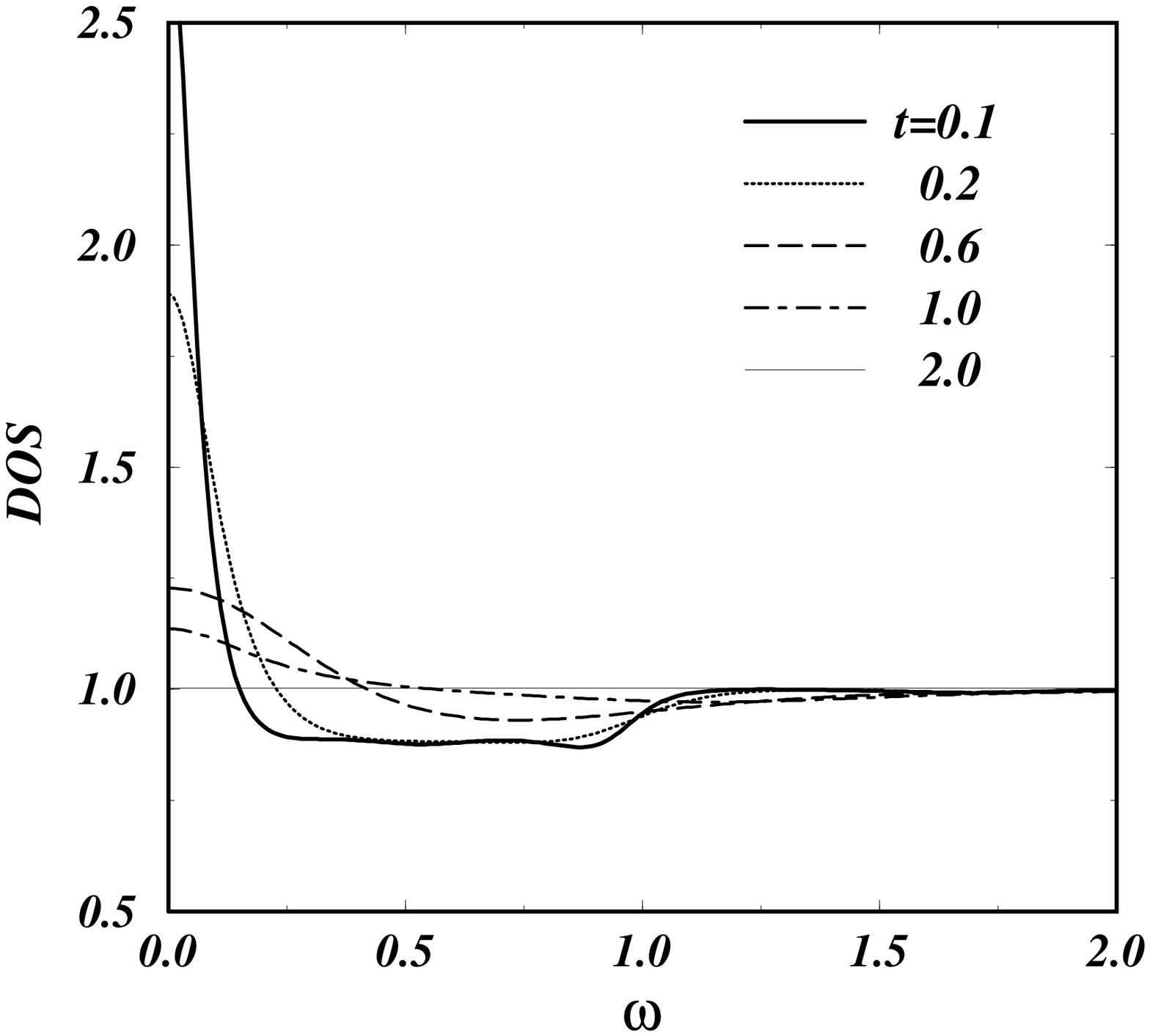, width=6.5cm,height=5cm}} 
%\centerline{\vbox{\hrule width 5cm height0.001pt}}
\vspace*{13pt}
\fcaption{- The density of states $N(\omega $ $)$ for the $FEP$ scattering with the given values of 
$t\equiv \pi T/\Omega$.}
\end{figure}
which narrows by lowering $T$, and pseudogap like behavior below $\Omega 
$. The pseudogap feature disappears at $T$ comparable with the phonon energy 
$\Omega $.

\section{Superconductivity due to FEP}
\noindent
%In further analysis it is
%assumed: 1) that the pairing is due to the forward electron-phonon
%scattering - the $FEP$ pairing; 2) the E-P interaction is considered in
%the weak coupling limit; 3) the effects of the $FIS$ impurities on $T_c$
%are studied in the self-consistent Born approximation.

%\subsection{$T_c$ due to FEP pairing in clean systems}
%\noindent
 In the weak
coupling limit 
 one obtains $Z(\vec k,n)=1$ and the $FEP$ pairing
gives the maximum $T_c$ on the Fermi surface, i.e. for $\xi =0$ where $\bar 
\xi _n(\vec k,n)=0$ . The solution of Eqs.(1) in the weak coupling limit
and for $T_c\ll \Omega $ is given by $T_{c0}=\lambda /4N(0)$, where $\lambda
=N(0)V_{ep}$.
Several points should be stressed: first, $T_{c0}$ is linear in $\lambda $
in the case of  $FEP$ pairing, similar to Ref.3.  
Second, in the
case of the electron-phonon pairing potential $V_{ep}(\vec q,\omega _n)$
with a finite cut-off ($q_c\neq 0$) it is shown in Ref.4 that $
T_{c0}\equiv T_{c0}(q_c=0)$ is  the value of $T_c$ zeroth-order in $q_c$. For
the short-range pairing potential ($q_c\sim 2k_F$), i.e. when $q_cV_F\sim
W\sim 1/N(0)$, one obtains the standard BCS result $T_{c0}^{BCS}=1.13\Omega
\exp (-1/\lambda )$, while for the long-range pairing potential ($q_cV_F\ll
\Omega $) the finite-$q_c$ correction to $T_{c0}$ is given by $T_c\simeq
T_{c0}\left( 1-7\zeta (3)q_cV_F/4\pi ^2T_{c0}\right) $.

%\subsection{$T_c$ due to $FEP$ pairing in the presence of $FIS$ impurities}
\noindent
 The FS impurities affect $T_c$, which is due to the FEP pairing. Some limiting cases are considered: (a) \underline{$\Gamma _F\ll \pi
T_c$} : One obtains $T_c=T_{c0}[1-4\Gamma _F/49T_{c0}]$; (b) \underline{$
\Gamma _F\gg \pi T_c$}: If $\Gamma _F\gg \Omega /2$ is fulfilled one
obtains $T_c\approx (\pi /2\gamma )\Omega \exp (-\pi \Gamma _F/V_{ep})$.
Note that the $FIS$ impurities are \underline{pair
weakening} for the $FEP$ pairing, i.e. there is an exponential fall-off of $
T_c$ with the increase of $\Gamma _F$.
One can show that the $NIS$ impurities
are pair weakening for $s-wave$- and pair breaking for $d-wave$ $FEP$ pairing\cite{DDKO}.

\section{Conclusions}
\noindent
 In summary, it is shown here that: $(a)$ the 
$FEP$ interaction and the $FIS$ scattering change the quasiparticle spectral
properties significantly. In the case of the $FEP$ interaction there are
pseudogap features in the density of states $N(\omega )$ for $\omega <\Omega 
$ (phonon frequency); $(b)$ by assuming that the pairing is due to the
forward E-P ($FEP$) scattering the critical temperature of clean systems $
T_{c0}$ depends linearly on the E-P coupling constant $\lambda $ in the
Migdal approximation;
$(c)$ the $FIS$ impurities affect in the same way the $s$- and $d$-$wave$ 
$FEP$ pairing and they are pair-weakening for both pairings; $(d)$ the $NIS$
isotropic impurities are pair-weakening for the $s$-$wave$ $FEP$ pairing and
pair-breaking for the $d$-$wave$ $FEP$ pairing.

\end{document}